# A Heliosheath Model for the Origin of the CMB Non-Monopole Spectrum

H.N. Sharpe

Bognor, Ontario, Canada
sh3149@brucetelecom.com

## ABSTRACT

A non-cosmological origin for the CMB low-order multipoles and associated anomalies is suggested in this paper. We discuss the possibility that these features of the power spectrum originate in the termination shock (TS) region of the heliosheath that surrounds the solar system. If the intrinsic CMB spectrum is assumed to be a pure monopole (2.73K) then thermal processes occurring within an ellipsoidal-shaped plasma region of the TS region could imprint the observed low-order multipoles and their alignment (so-called 'axis of exil") onto this background CMB. A key requirement of the model is that the TS plasma be characterized as an optically thin graybody with non-LTE perturbations to explain both the blackbody and non-blackbody anomalies. Non-thermal TS processes are also discussed. TS synchrotron radiation is shown as a possible cause for the reported ARCADE 2 CMB distortion. Magnetoacoustic oscillations and heliosheath turbulence are suggested as possible sources for the higher-order multipole moments of the CMB. Data from the ongoing Voyager missions is critical to this study. Research problems are identified for future work to better validate the heuristic models presented here.

Cosmic microwave background;Sun;96.50.Ek;98.70.Vc

## INTRODUCTION

It is generally accepted that the observed temperature fluctuation power spectrum of the cosmic microwave background (CMB) is in fact cosmological in origin. Extraordinary efforts have been taken to remove all possible sources of contamination. Interpretation of the observed CMB fluctuations has played a key role in the development of the concordant cold dark matter model ($\Lambda$ CDM) for the evolution of the Universe [1].

However an important source of contamination may have been overlooked. Although originally believed to be perfectly spherical, recent results from the two Voyager spacecraft indicate that the solar wind termination shock (TS) which surrounds our solar system is asymmetric [21]. The TS forms where the outflowing supersonic solar wind is slowed to subsonic through its interaction with the interstellar wind. It marks the inner boundary of the heliosheath. The heliosheath is characterized by a turbulent, magnetized plasma. The outer boundary is the heliopause and occurs where the solar wind pressure balances that of the interstellar medium.

Analysis of the Voyager data collected while these spacecraft crossed the TS into the heliosheath shows the TS to be a boundary of abrupt changes in pressure, temperature, density, magnetic and electric field properties and ion and electron (plasma) properties [3,8]. It should therefore be expected that the TS (and heliosheath) will leave an imprint on observed radiation which originates outside of the solar system and propagates through this optically thin boundary. We suggest that this imprint could be the low-order multipole moments of the CMB power spectrum and possibly the entire spectrum as well.

For this analysis the CMB is assumed to be perfectly isotropic outside the heliosphere. In a sense the TS may introduce an "astigmatism" into the CMB.

Termination shock physical processes can interact with the CMB radiation in two general ways. The CMB photons can be absorbed and scattered by local radiative and kinetic processes as they propagate through the TS region. Or they can pass unaffected through an essentially transparent heliosheath. In this paper we focus primarily on the latter case. We assume that local thermal and non-thermal processes in the TS emit radiation that is simply superimposed on the background CMB. Since this radiation is emitted on the TS surface it should acquire the geometric properties of the TS. Accordingly the observed CMB spectrum inside the solar system could reflect the multipole distortions of the TS.

We begin this paper with a discussion of the geometric model for the TS surface. We demonstrate that an ellipsoid of revolution could mimic the observed CMB quadrupole. Ellipsoidal distortions have been previously used to explain the CMB quadrupole but in those cases it was the Universe that was distorted rather than the TS surface [22]. We also suggest that a more generally distorted ellipsoid could explain additional anomalies in the low-order multipole CMB power spectrum. These include the apparent alignment of the quadrupole and the octupole ($\ell=2,3$) moments (and possibly $\ell=4$ and 5) with the direction of the solar system's motion and the ecliptic plane (the so-called "axis of evil"), as well as the low power in $\ell=2,3$ [23].

We next proceed to discuss physical mechanisms which could be operating in the TS surface region. Both thermal and non-thermal radiation are considered. A key requirement for thermal radiation is that there exist at least one identifiable velocity population in local thermodynamic equilibrium (LTE). If the TS region can be characterized as an optically thin graybody with non-Maxwellian perturbation distributions, then both the blackbody and non-blackbody low-order CMB multipole anomalies could be explained with this model. Non-thermal processes include synchrotron radiation generated by energetic electrons spiraling around the heliosheath magnetic fields. We show how this radiation could explain the ARCADE 2 CMB distortions.

The model presented in this paper is heuristic. Several major areas of research are identified at the end to better validate this model. Given the potential for heliosheath processes to "contaminate" an otherwise perfectly isotropic 2.73K CMB monopole it is hoped that some of these problems will be studied.

## GEOMETRIC MODEL

We represent the TS by an idealized prolate ellipsoid of revolution; a:a:c where c > a. The equation for the surface is:

$$r(\theta,\phi) = a^2c[\, a^2c^2\sin^2\theta + a^4\cos^2\theta\,]^{-1/2} \tag{1}$$

The usual spherical harmonic expansion expresses (1) as :

$$r(\theta,\phi) = \sum_{n=0}^{\infty}\sum_{m=-n}^{n} C_{nm}\, Y_n^m(\theta,\phi) \tag{2}$$

The rms amplitude for the nth moment is:

$$r_n = \left[\frac{1}{4\pi}\sum_{m=-n}^{n}C_{nm}^2\right]^{1/2} \qquad (3)$$

From symmetry only those coefficients with m=0 are nonzero. Also, because (1) is even in $\theta$, there are no moments with odd n such as the dipole and octupole. We then derive the following expression for the ratio of the quadrupole moment to the monopole moment for this surface:

$$\frac{r_2}{r_0} = \frac{C_{20}}{C_{00}} = \frac{3\sqrt{5}}{4}\left[\frac{1}{f^2} - 2\left(\frac{1}{3} + \frac{1}{g\sqrt{\left(\frac{c}{a}\right)^2 - 1}}\right)\right] \qquad (4)$$

where: $f \equiv \sqrt{1-\left(\frac{a}{c}\right)^2}$ and $g \equiv \pi - 2\cos^{-1}f$

The CMB temperature anisotropies may also be represented by (3), where $r_n$ is replaced by $T_n$. The observed constraint on the rms temperature ratio of the quadrupole to monopole moments is [1,2]:

$$\frac{T_2}{T_0} \approx 5x10^{-6} \qquad (5)$$

In the CMB spherical harmonic analysis all the $C_{nm}$ in (3) are generally nonzero. However in the present simple model we wish to determine the ratio (a/c) in (4) which accounts for this observed CMB quadrupole. Equating (4) and (5) then we find that (a/c) = 0.95.

Richardson et al [3] report a N-S asymmetry for the TS of 7-8 AU with a variability of 2-3 AU due to solar wind pressure changes. This distortion is comparable to that just computed for the prolate ellipsoid assuming an effective mean radius of 90 AU for the TS. Clearly these results are not conclusive as the TS is not an idealized ellipsoidal surface with zero dipole moment. They do suggest however the possibility that the CMB quadrupole moment may be imprinted by the geometric distortion of the TS.

To include the odd-numbered harmonics a more generalized ellipsoidal distortion to the TS must be considered:

$$\frac{x^2}{a^2} + \frac{y^2}{b^2} + \frac{z^2}{c^2} = 1 \qquad (6)$$

where a,b,c, are chosen to satisfy the multipole constraints on $T_\ell/T_0$, $\ell=2,3,4$ and possibly $\ell=5$. This is an optimization problem. The result will be an ellipsoidal surface which satisfies the rms magnitudes of multipoles 2 to 5 and their alignment with the Sun's motion and the ecliptic plane since the principal distortion of the TS is in the Sun's direction through the ISM. It should also be noted that this ellipsoid will generate a dipole moment ($\ell$=1). This geometric moment should however be subsumed in the much larger Doppler dipole, though its value may have physical implications for the inferred peculiar velocity of the Local Group. With sufficient computing power a more generalized and rigorous treatment of this problem would involve a full spherical harmonic expansion of the ellipsoidal surface with a constrained optimization of the $C_{nm}$ for the observed multipoles $\ell$=2 to 5. Most generally, the full CMB thermal power spectrum for multipoles greater than zero could be represented in the TS surface distortion relative to the sphere.

## PHYSICAL MECHANISMS

### Thermal Heliosheath Radiation

Termination shock physical processes can interact with the CMB radiation in two general ways. The CMB photons/waves can be absorbed, scattered and refracted by local radiative, kinetic and optical (geometric) processes as they propagate through the TS region. Or they can pass unaffected through an essentially transparent heliosheath. In this section we focus on the latter case. We assume that local thermal processes in the TS emit radiation that is simply superimposed on the background CMB. Since this radiation is emitted on the TS surface it should acquire the geometric properties of the TS. Accordingly, the observed CMB spectrum inside the solar system will reflect the multipole distortions of the TS.

Start with the general transfer equation for a pencil of radiation propagating through an optically thin region characterized by optical depth $\tau_\nu$ and source function $S_\nu$:

$$I_\nu(\tau_\nu) = I_\nu(0) + \tau_\nu S_\nu \qquad \tau_\nu \ll 1 \qquad (7)$$

Identify $I_\nu(0)$ with the CMB background brightness spectrum and $\tau_\nu S_\nu$ with the additional brightness from the TS region. For the geometric distortions of the TS to manifest as blackbody multipoles against an otherwise isotropic CMB background, the physical mechanism operating in the TS region must be in local thermodynamic equilibrium (LTE). First we demonstrate how this process could work and then discuss its relevance.

In equation (7) we make the following key assumptions:

1. The source function is Planckian (LTE): $S_\nu = B_\nu$
2. The absorption coefficient (opacity) may be represented as a graybody (independent of frequency) with a small departure from grayness, $\delta_\nu(\theta)$ that in general could depend on direction:

$$\tau_\nu \rightarrow \bar{\tau}(1+\delta_\nu(\theta)) \qquad \bar{\tau} \equiv \bar{\alpha}L \ll 1 \text{ and } \delta_\nu \ll 1$$

Assumption 2 preserves the blackbody profile of the emitted thermal radiation while allowing for small deviations. A graybody model for the TS plasma may be justified by noting that the interplanetary dust is constantly being swept in to the heliosheath by the solar wind. The resulting "dusty plasma" may show graybody properties down to radio frequencies depending on the dust properties [4,5]. This is a topic for further study. With these assumptions and using Rayleigh-Jeans equation (7) becomes:

$$T_v^{obs} = T_{cmb} + \bar{\tau}T^{TS} + \bar{\tau}T^{TS}\delta_v(\theta) \tag{8}$$

The first term on the rhs is the CMB monopole 2.73K. The second term is the graybody Planckian perturbation to the CMB monopole caused by the TS geometric distortion. It may be thought of as a normalization constant. $T^{TS}$ is the effective local kinetic temperature in the TS and $\bar{\tau}T^{TS}$ is the observed TS brightness temperature.

The third term on the rhs is the small departure from a blackbody spectrum due to non-grayness of the thermal radiation. If we assume it has no directional dependence on the TS surface then this non-blackbody contribution will inherit the same geometric distortion as the blackbody perturbations. This could be a cause for the observed alignment of the non-blackbody multipoles with the overall CMB multipoles [6,7].

The assumption of graybody thermal radiation is key to this model. It requires that detailed balance holds for collisional processes in the TS region, which in turn requires that the particle distribution function is Maxwellian [5]. Under these conditions a local kinetic temperature can be defined and LTE is valid.

Since the TS is a shock interface between the supersonic solar wind and the heliosheath region, the shock can disturb the tendency towards LTE. Several recent reports discuss the findings of the Voyager spacecraft which have now penetrated the TS region [3,8,9]. They show a chaotic, turbulent magnetized plasma far from LTE. Nevertheless, it may still be possible to model this region globally as a perturbed Maxwellian graybody distribution. On-going research into this difficult problem is continuing. *If an appropriate distribution function for the TS can be developed it will provide an important local alternative explanation to cosmic models for the observed low-order CMB anomalies*. In addition the TS model can continue to be refined with more data as the Voyager spacecraft penetrate deeper into the heliosheath.

Low Quadrupole "Anomaly" Model

In the model just presented we assumed an isotropic monopole for the CMB background radiation. All observed blackbody and non-blackbody low-order deformations were attributed to geometric distortions of the TS and in situ LTE processes ( with small departures from Planckian). In [6] Diego et al assume a cosmological origin of the full CMB multipole spectrum, but attempt to explain the low quadrupole power ( low relative to inflation model predictions) and its alignment with the octupole and the ecliptic plane in terms of potential sources of contamination. They present a quasi-blackbody "toy" model normalized such that its non-blackbody amplitude is approximately 10 $\mu$ K in the V+W-2Q band (after filtering with a 7 deg Gaussian). They then demonstrate that this "toy" model possesses a quadrupole which is anti-correlated with the WMAP5 quadrupole. When subtracted from the observed quadrupole the result better approximates the value expected from theory. The octupole is not affected by their model since significant power is only found in the even moments. The low order multipoles also seem to come into better alignment.

While this "toy" model is capable of explaining the observed anomaly, Diego et al acknowledge that a physical justification for its existence is lacking. We conclude this section with a possible justification for their "toy" model.

We assume here the simple ellipsoid of revolution discussed earlier for the TS distortion. Then we may write equation (7) as:

$$I_\nu^{obs} = I_\nu(0) + B_\nu[\bar{\tau}(1+\delta_\nu)] \tag{9}$$

where $I_\nu(0)$ is the full CMB multipole background and the second term represents the quasi-blackbody distortion due to non-LTE processes on the TS as discussed above. The normalization can be chosen such that this term satisfies the same $10\,\mu K$ constraint in the V+W-2Q band. Since the distortion originates on the TS it will share its geometric properties of only possessing even multipoles (quadrupole). Hence equation (9) could be a physical justification of the "toy" model proposed by Diego et al.

### Non-thermal Heliosheath Radiation Sources

Seiffert et al [10] have presented data from the ARCADE 2 instrument that suggests a residual signal beyond the CMB background after correcting for potential contributions from various extra-galactic sources. The spectrum function they fit to the ARCADE (and other) data has the following form:

$$T(\nu) = T_0 + A\,(\nu/1\text{GHz})^\beta + \Delta T(\nu) \tag{10}$$

where $T_0$ is the CMB baseline temperature, A is the power law amplitude at 1 GHz, $\beta$ is the power law index, $\nu$ is the frequency and $\Delta T(\nu)$ is a CMB spectral distortion. They find that the unexplained residual emission is consistent with a power law amplitude $A = 1.06 \pm 0.11$K at 1 GHz and a spectral index $\beta = -2.56 \pm 0.04$.

In this section we suggest that the residual emission may in fact originate from an optically thin layer within the heliosheath which surrounds the solar system. The energetic electrons follow a power law spectrum in the presence of random magnetic fields. In this idealized model, the synchrotron radiation is homogeneous and isotropic.

We re-write equation (7) in terms of the thickness L for a portion of the optically thin heliosheath (negligible absorption coefficient),

$$I_{obs} = I_{source} + L\,j_{shell} \tag{11}$$

where $j_{shell}$ is the specific emissivity for the radiating shell. Next we assume a power law energy distribution for the radiating electrons with spectral index p:

$$N(E)dE = K\,E^{-p}\,dE \tag{12}$$

Then the observed brightness is:

$$I_{obs} = I_{source} + 1.35\times10^{-22}\,K\,B^{(p+1)/2}\,a(p)L\,(6.26\times10^{18}/\nu)^{(p-1)/2} \tag{13}$$

where B is the magnetic field and a(p) is a function used to simplify the power law index dependency [11]. The units of I are $\text{erg cm}^{-2}\,\text{sec}^{-1}\,\text{ster}^{-1}\,\text{Hz}^{-1}$. K has units $\text{ergs}^{p-1}\text{cm}^{-3}$.

We convert to the observed brightness temperature using the Rayleigh-Jeans relation: $T_{obs}=(c^2/2k\nu^2)I_\nu$ where k is the Boltzmann constant:

$$T_{obs} = T_{source} + [\,4.4\times10^{14}\,\text{K L a(p) B}^{(p+1)/2}\,(6.26\times10^{18})^{(p-1)/2}\,]\,\nu^{-(p+3)/2} \quad (14)$$

At this point we identify $T_{source}$ with $T_0 + \Delta T(\nu)$ in (10) and focus on the temperature spectrum contribution from the optically thin intervening shell.

The power law index $\beta$ from the ARCADE data fit implies a spectral index for the radiating electrons of p = +2 (from (p+3)/2 = 2.5). A value of +2 is typical for relativistic electrons assuming an isotropic pitch angle distribution. The corresponding value for a(p) is 0.103 [11].

Using these values in (14) for the shell brightness temperature spectrum we obtain an observed brightness temperature for the shell ($\nu$ in GHz):

$$T_{shell} = [\,3.48\,\text{K L B}^{3/2}\,]\,\nu_{GHz}^{-2.5} \quad (15)$$

Integrating (5) and using p = 2, K is seen to be an electron energy density for this case:

$$K = n_0 E_1 \quad (16)$$

where $n_0$ is the electron number density in the shell and $E_1$ is the lower bound on the energy interval. We identify $E_1$ with the critical electron energy $E_{cr}$, associated with the critical frequency. Finally we have the desired functional form that corresponds to the residual emission in (10):

$$T_{shell} = A\,\nu_{GHz}^{-2.5} \quad (17)$$

where $A \equiv 3.48\,n_0\,E_{cr}\,L\,B^{3/2}$. We wish to investigate the conditions under which the value of A equals the fitted power law amplitude value of 1 Kdeg in (10). The four physical factors which contribute to A are the electron number density, the electron energy that corresponds to the critical frequency, the magnetic field and the shell thickness. As a first pass we compute the required value of L given representative values of the heliosheath parameters. Using $n_0 = 0.01\,\text{cm}^{-3}$, $B = 1\,\mu G$ [3,12] and $E_{cr} = 1$ Gev which corresponds to a critical frequency of 16MHz, we compute L ~ 10 AU for A = 1Kdeg. The heliosheath is expected to be of this order thickness [13]. The essential point is that the required thickness of the optically thin shell is not inconsistent with the power law amplitude of 1K deg.

Next we consider the nature of the heliosheath magnetic fields and the energetic electrons. Alfven [14] developed a heliospheric current model along the lines of the observed terrestrial magnetospheric current system. Briefly, the Sun acts as a unipolar inductor that generates an emf which drives two polar current systems. These currents spread out, possibly to the heliosheath, and return to the Sun near the equatorial plane thereby completing the "circuit". It is not known if the polar currents spread out to form current sheets or if they concentrate into filaments. It is likely that they break up into cylindrical filaments and pinch down in a Bennett-type pinch effect. But whereas the Bennett pinch represents a balance between the plasma pressure and a compressing electromagnetic force, a low density plasma like that in the heliosheath cannot generate an internal plasma pressure. The result is a force-free magnetic field in a reference frame which moves with the current because the electric and magnetic fields are aligned in this frame. To an outside observer the magnetic field would appear as a twisted magnetic "rope" [15]. Force-free fields represent the lowest state of magnetic energy that a closed system may attain [16]. Hence an initial current sheet should break into Birkeland (magnetic field aligned) vortex tubes having a characteristic distribution function. Recent Voyager data suggest the presence of magnetic rope structures in the heliosheath [17].

Relativistic electrons in the heliosheath could be produced in a system of double layers distributed along the field-aligned current systems. The double layer is an electrostatic structure a few Debye lengths wide which may appear in a current carrying plasma. It can sustain a high net potential difference which could accelerate electrons traversing it. It may then provide a mechanism for transforming stored magnetic energy into the directed kinetic energy of the accelerated particles. Alfven [18] discusses the production of relativistic electrons in double layers associated with a heliospheric current system similar to the mechanism he proposed for the Earth's magnetosphere. Intense localized plasma-wave electric fields recently observed at the TS [8] could accelerate electrons to these relativistic energies.

Validation of this mechanism is difficult. It must be shown that a power spectral index of +2 can actually be derived for the radiating electrons in a distribution of magnetized vortex tubes. It must also be shown that the resultant radiation is isotropic, though it may contain higher moments due to distortion of the heliosheath by the local ISM. Finally improved observational and theoretical constraints are required for the parameters that comprise the power law amplitude in (17). Overlying all these problems is the need to better understand heliosheath physics and whether the heliosheath can in fact support a global current system. A correlation between the residual emission and the solar cycle could suggest this support.

**Other Mechanisms**

Radiation scattering mechanisms like Thomson, Compton and inverse-Compton scattering are not expected to be important in the heliosheath due to the low ion/electron concentrations and the low temperature. The optical depth for electron scattering is essentially zero so scattering of both CMB photons and in situ radiation should not be important. But note that dust could be a factor as discussed earlier.

Thermal bremsstrahlung ( free-free) emission is also negligible in the TS/heliosheath again due to low ion/electron concentrations.

Radio wave refraction of the CMB in the TS region, due to a change in the index of refraction could also imprint the TS geometric distortion on the observed CMB spectrum. Since ($I_v/n_r^2$) is constant along a CMB ray through the TS, ($I_v$ is the specific intensity and $n_r$ is the index of refraction) the CMB would suffer a distortion which mimics the shape of the TS. At WMAP frequencies this effect is negligible if one only considers the electron density (small plasma frequency). But the observed sharp increases in the temperature, plasma density and magnetic field strength across the TS [3] may produce an additional effect on the refractive index which needs to be studied.

The observed small-scale power spectrum of the CMB anisotropies could also have its origin from in situ dynamical TS processes. This power spectrum is usually interpreted in terms of primordial fluctuations and is seen as confirmation of the concordance model of cosmology. But we may in fact be viewing the cosmos through an optically thin "dirty window". The heliosheath plasma is modeled with MHD codes [19,20] which have the possibility of developing magnetoacoustic waves. These waves could mimic the high order multipoles seen in the CMB power spectrum. Also the chaotic turbulent nature of this surface region could leave its imprint on the CMB spectrum, including polarization.

Finally we mention that the TS is expected to change shape with the solar wind pressure [3]. Any observed temporal changes of the CMB power spectrum would then be a good test of the model proposed in this paper. Recently temporal low harmonic residuals have been reported from a comparison of the WMAP3 and WMAP5 maps [2] though they are attributed to re-calibration effects.

## DISCUSSION

The Voyager spacecraft have provided a unique opportunity to characterize the dynamical and radiative processes in the heliosheath which surrounds the solar system. Since the CMB is viewed through this "window" it should be expected that these processes may leave their imprint on the CMB spectrum. This paper has identified possible sources of heliosheath/TS contamination and the role they may play in addressing some of the observed anomalies. However a much more rigorous examination of these physical processes is required given their potential to impact the concordance model of cosmology. We outline below some of the major research problems.

1. A rigorous thermodynamical model is required for the heliosheath plasma and the termination shock region. If a dusty plasma model can be developed which characterizes this region as an optically thin graybody with non-Maxwellian perturbation distributions, then most of the low-order multipole features of the CMB spectrum could be explained by a local rather than cosmic origin. In addition, non-blackbody spectrum anomalies could be explained by non-LTE velocity populations.

2. The high order multipole components should be studied with a full MHD turbulence model for the TS region to determine if the observed acoustic peaks can also be given a local interpretation.

3. The role of heliosheath electric and magnetic fields in the observed polarization of the background CMB should be studied. In situ generated synchrotron radiation, which has been identified in this paper as a possible source of the reported Arcade 2 anomaly should also be studied for polarization effects on the CMB. The mechanism for the production of relativistic electrons must also be identified.